\documentclass[12pt,a4paper]{article}
\usepackage{amsmath,amsfonts,amsthm,graphicx,lscape}
\usepackage[dvips]{psfrag}
\numberwithin{equation}{section}

\def\beqa{\begin{eqnarray}}
\def\enqa{\end{eqnarray}}
\def\beq{\begin{equation}}
\def\enq{\end{equation}}

\begin{document}
\title{A note on the 
 high temperature expansion of the density matrix for 
 the isotropic Heisenberg chain 
}
\author{Zengo Tsuboi 
\footnote{
E-mail address: 
zengo\_tsuboi@pref.okayama.jp}
\\
{\it Okayama Institute for Quantum Physics
\footnote{URL: 
http://www.pref.okayama.jp/kikaku/kouryoushi/english/kouryoushi.htm
},}
 \\
{\it Kyoyama 1-9-1, Okayama 700-0015, Japan}
}
\date{}
\maketitle
\begin{abstract} 
G\"ohmann, Kl\"umper and Seel derived 
the multiple integral formula of 
the density matrix of the 
$XXZ$ Heisenberg chain at finite temperatures. 
We have applied the high temperature expansion (HTE)
method to isotropic case of their formula in  
a finite magnetic field and 
obtained coefficients for several short-range
 correlation functions. 
For example, we have succeeded to obtain 
the coefficients of the HTE of the 3rd neighbor correlation function 
$<\sigma_{j}^{z}\sigma_{j+3}^{z}>$
for zero magnetic field up to the order of 25. 
These results expand our previous results on the emptiness formation probability 
[Z.Tsuboi, M.Shiroishi, J. Phys.A: Math. Gen. 38(2005) L363] 
to more general correlation functions. 
\end{abstract}
 \\
{\it Report-no:} {\bf OIQP-06-10}  \\
{\bf to appear in Physica A}
\section{Introduction} 
G\"ohmann, Kl\"umper and Seel derived \cite{GKS04-3} 
(see also \cite{GKS04-1,GKS04-2,GS05,GHS05})
 a multiple integral formula of matrix elements of a 
density matrix of a finite segment of arbitrary length $m$ 
of the  anti-ferromagnetic spin $1/2$ $XXZ$ Heisenberg infinite chain at finite 
temperature in a finite magnetic field 
by combining the quantum transfer matrix approach \cite{S85}-\cite{S03} and 
the algebraic Bethe ansatz technique. 
Their formula generalizes the multiple integral formulae for zero temperature 
\cite{JMMN92,JM96,KMT00} to finite temperature case. 
This is a fundamental quantity since thermal average of any 
operators acting non-trivially on the segment of length $m$ can be expressed 
in terms of their formula. 
Thus it is an important problem to perform this multiple integral and 
extract concrete numbers from it. 
Their formula contains an auxiliary function, which is a solution 
of a nonlinear integral equation. This nonlinear integral equation is 
 essentially same as the one for the free energy \cite{K92,DD92}. 
Thus to evaluate their formula consists of two non-trivial tasks:
 to solve the nonlinear integral equation and to integrate the multiple integrals. 
In our previous paper \cite{TSh05},
 we applied the high temperature expansion (HTE) method to 
 a multiple integral formula \cite{GKS04-2} 
of the emptiness formation probability $P(m)$ for the $XXX$ model,
 which is the probability of $m$ adjacent spins being aligned upward, 
and succeeded to obtain the coefficients of $P(3)$ up to the order of 42.   
As for zero magnetic field case, there is also numerical calculation \cite{BG05} 
 for the multiple integral for $m=2,3$. 
The purpose of this paper is to expand our previous results on the HTE for 
$P(m)$ \cite{TSh05} to more general correlation functions for 
the spin $1/2$ isotropic Heisenberg chain in 
a magnetic field $h$. 
In section 2, we introduce the  multiple integral formula of the 
matrix elements of the density matrix \cite{GKS04-3}. 
In section 3, we evaluate this multiple integral by 
the HTE method. Based on the result of the HTE of the density matrix,
 we will calculate the HTE of 
two point correlation functions (\ref{2ndzz})-(\ref{2ndpm}). 
In particular for zero magnetic field case, we have succeeded to 
obtain the coefficients of the HTE of a 3rd neighbor correlation 
function up to the order of 25 (cf. eq. (\ref{3rdh=0})). 
Section 4 is devoted to concluding remarks. 
\section{Integral representation of the density matrix}
The Hamiltonian of the spin-$1/2$ isotropic Heisenberg chain in 
a magnetic field $h$ is given as 
\begin{eqnarray}
H=J\sum_{j=1}^{L}(\sigma_{j}^{x}\sigma_{j+1}^{x}+
\sigma_{j}^{y}\sigma_{j+1}^{y}
+\sigma_{j}^{z}\sigma_{j+1}^{z})-\frac{h}{2}\sum_{j=1}^{L}\sigma_{j}^{z},
\label{hamil}
\end{eqnarray}
where $\sigma_{j}^{x}$, $\sigma_{j}^{y}$, $\sigma_{j}^{z}$  
are the Pauli matrices which act non-trivially on the $j$-th lattice site in 
a chain of length $L$. 
Here the periodic boundary condition 
$\sigma_{j+L}^{k}=\sigma_{j}^{k}$ is assumed. 

Let us introduce $2\times 2$ matrices: 
\begin{eqnarray}
e^{1}_{1}=
\left(
\begin{array}{cc}
1 & 0 \\
0 & 0
\end{array}
\right)
, \quad
e^{2}_{1}=
\left(
\begin{array}{cc}
0 & 1  \\
0 & 0
\end{array}
\right)
, \quad
e^{1}_{2}=
\left(
\begin{array}{cc}
0 & 0 \\
1 & 0
\end{array}
\right)
, \quad
e^{2}_{2}=
\left(
\begin{array}{cc}
0 & 0 \\
0 & 1
\end{array}
\right)
\end{eqnarray}
These matrices are embedded into the space $({\mathbb C^{2}})^{\otimes L}$ 
on which the Hamiltonian (\ref{hamil}) acts: 
\begin{eqnarray}
{e_{j}}^{\alpha}_{\beta}=I^{\otimes (j-1)}\otimes 
e^{\alpha}_{\beta} \otimes I^{\otimes (L-j)}, 
\end{eqnarray}
where $I=e^{1}_{1}+e^{2}_{2}$,  $\alpha,\beta \in \{1,2\}$ and $j \in \{1,2,\dots,L \}$.
The above Pauli matrices can be written in terms of these matrices: 
$\sigma^{x}_{j}={e_{j}}^{1}_{2}+{e_{j}}^{2}_{1}$, 
$\sigma^{y}_{j}=i{e_{j}}^{1}_{2}-i{e_{j}}^{2}_{1}$, 
$\sigma^{z}_{j}={e_{j}}^{1}_{1}-{e_{j}}^{2}_{2}$. 
We also put $\sigma^{+}_{j}={e_{j}}^{2}_{1}$ and 
$\sigma^{-}_{j}={e_{j}}^{1}_{2}$. 

G\"ohmann, Kl\"umper and Seel 
derived \cite{GKS04-3} an integral representation of the 
density matrix of the $XXZ$ chain at finite temperature $T$. 
The isotropic ($XXX$) limit of their formula is given as follows. 
\begin{eqnarray}
<{e_{1}}^{\alpha_{1}}_{\beta_{1}}
{e_{2}}^{\alpha_{2}}_{\beta_{2}}
\cdots 
{e_{m}}^{\alpha_{m}}_{\beta_{m}}>
&=& \lim_{L \to \infty} 
\frac{{\mathrm Tr} {e_{1}}^{\alpha_{1}}_{\beta_{1}}
{e_{2}}^{\alpha_{2}}_{\beta_{2}}
\cdots 
{e_{m}}^{\alpha_{m}}_{\beta_{m}} e^{-\frac{H}{T}}}{{\mathrm Tr} e^{-\frac{H}{T}}}
\nonumber \\ 
&=&
\prod_{j=1}^{|\alpha^{+}|}\int_{C}\frac{dy_{j}}{2\pi (1+\mathfrak{a}(y_{j}))} 
(y_{j}-i)^{{\widetilde \alpha}_{j}^{+}-1}
y_{j}^{m-{\widetilde \alpha}_{j}^{+}}
\nonumber \\ 
&& \hspace{-80pt} \times
\prod_{j=|\alpha^{+}|+1}^{m}\int_{C}\frac{dy_{j}}
{2\pi (1+\overline{\mathfrak{a}}(y_{j}))} 
(y_{j}+i)^{{\widetilde \beta}_{j}^{-}-1}
y_{j}^{m-{\widetilde \beta}_{j}^{-}}
\nonumber \\ 
&&  \hspace{-80pt} \times 
\det_{1\le j,k \le m}
 \left(\frac{\partial_{\xi}^{(k-1)} G(y_{j},\xi)|_{\xi=0}}{(k-1)!} \right)
 \frac{1}
      {\prod_{1 \le j < k \le m}(y_{j}-y_{k}-i)},
      \label{density}
\end{eqnarray}
where 
$\mathfrak{a}(v)$ and $G(v,\xi)$ are solutions of the following integral 
equations. 
\begin{eqnarray}
\log \mathfrak{a}(v)=-\frac{h}{T}+\frac{2J}{v(v+i)T} - \int_{C}\frac{dy}{\pi}
 \frac{\log (1+\mathfrak{a}(y))}{1+(v-y)^{2}},
  \label{nlie1}
\end{eqnarray}
\begin{eqnarray}
G(v,\xi)= -\frac{1}{(v-\xi )(v-\xi-i)} + \int_{C}\frac{dy}{\pi}
 \frac{1}{1+(v-y)^{2}} \frac{G(y,\xi)}{1+\mathfrak{a}(y)}.
 \label{nlie2}
\end{eqnarray}
Here $(\alpha_{n})_{n=1}^{m}$ and $(\beta_{n})_{n=1}^{m}$ are sequences of 
$1$ or $2$. We define the number of $1$ in $(\alpha_{n})_{n=1}^{m}$ 
as $|\alpha^{+}|$ and 
the position $n$ of $j$-th $1$ in $(\alpha_{n})_{n=1}^{m}$ as $\alpha_{j}^{+}$: 
$\alpha_{\alpha_{j}^{+}}=1$, 
$1 \le \alpha_{1}^{+}<\alpha_{2}^{+}<\cdots <\alpha_{|\alpha^{+}|}^{+} \le m$. 
We also define the number of 
$2$ in $(\beta_{n})_{n=1}^{m}$ as $|\beta^{-}|$ and 
the position $n$ of $j$-th $2$ in $(\beta_{n})_{n=1}^{m}$ as $\beta_{j}^{-}$: 
$\beta_{\beta_{j}^{-}}=2$, 
$1 \le \beta_{1}^{-}<\beta_{2}^{-}<\cdots <\beta_{|\beta^{-}|}^{-} \le m$. 
We shall put 
$\widetilde{\alpha}_{j}^{+}=\alpha_{|\alpha^{+}|-j+1}^{+}$ for 
$j \in \{1,2,\dots,|\alpha^{+}| \}$ and 
$\widetilde{\beta}_{j}^{-}=\beta_{j-|\alpha^{+}|}^{-}$ for 
$j \in \{|\alpha^{+}|+1,|\alpha^{+}|+2,\dots,|\alpha^{+}|+|\beta^{-}| \}$. 
The contour $C$ surrounds the real axis anti-clockwise manner. 
$\overline{\mathfrak{a}}(v)$ is defined as 
$\overline{\mathfrak{a}}(v)=1/\mathfrak{a}(v)$. 
The emptiness formation probability 
is a special case of this density matrix element:
$P(m)=<{e_{j}}^{1}_{1} {e_{j+1}}^{1}_{1} \cdots {e_{j+m-1}}^{1}_{1}>$. 
In this case, the multiple integral formula (\ref{density}) reduces to 
the one in \cite{GKS04-2}. 
\section{High temperature expansion}
In this paper, we will calculate the HTE of the 
following two-point correlation functions for finite magnetic 
field $h$: 
\begin{eqnarray}
&&  <\sigma_{j}^{z}\sigma_{j+2}^{z}>=
<({e_{j}}^{1}_{1}-{e_{j}}^{2}_{2})({e_{j+1}}^{1}_{1}+{e_{j+1}}^{2}_{2})
({e_{j+2}}^{1}_{1}-{e_{j+2}}^{2}_{2})>, \label{2ndzz} \\
&& <\sigma_{j}^{z}\sigma_{j+3}^{z}> \nonumber \\ 
&& \hspace{10pt}
=<({e_{j}}^{1}_{1}-{e_{j}}^{2}_{2})({e_{j+1}}^{1}_{1}+{e_{j+1}}^{2}_{2})
({e_{j+2}}^{1}_{1}+{e_{j+2}}^{2}_{2})
({e_{j+3}}^{1}_{1}-{e_{j+3}}^{2}_{2})>, \\
&& <\sigma_{j}^{+}\sigma_{j+2}^{-}>=
<{e_{j}}^{2}_{1}({e_{j+1}}^{1}_{1}+{e_{j+1}}^{2}_{2})
{e_{j+2}}^{1}_{2}>. \label{2ndpm}
\end{eqnarray}
At $h=0$, one can express nearest neighbor and 2nd neighbor correlation functions 
in terms of $P(2),P(3)$: 
$<\sigma_{j}^{z}\sigma_{j+1}^{z}>=4P(2)-1$, 
$<\sigma_{j}^{z}\sigma_{j+2}^{z}>=8(P(3)-P(2)+\frac{1}{8})$.  
Then we can immediately calculate the HTE of these correlation functions up to 
the order of 42 from the results in \cite{TSh05}. 
We can also calculate the HTE of $<\sigma_{j}^{+}\sigma_{j+2}^{-}>$  
up to the order of 42 from the relation 
$<\sigma_{j}^{+}\sigma_{j+2}^{-}>=\frac{1}{2}<\sigma_{j}^{z}\sigma_{j+2}^{z}>$ 
which holds for isotropic model ($XXX$-model) at $h=0$. 
Moreover one can calculate the nearest neighbor correlation functions 
by taking the derivative of the free energy of the $XXZ$-chain with respect 
to the anisotropy parameter. 
On the other hand, $<\sigma_{j}^{z}\sigma_{j+3}^{z}>$ can not be expressed 
only in terms of $P(m)$ even at $h=0$.  

At first, we will calculate the HTE of $\mathfrak{a}(v)$ from the NLIE (\ref{nlie1}). 
Note that this calculation is similar to the one for the 
HTE for the free energy \cite{DD92}. 
We assume the following expansion for small $|J/T|$:
\begin{eqnarray}
&& \mathfrak{a}(v)=\exp\left(\sum_{k=1}^{\infty}a_{k}(v)\left( \frac{J}{T} \right)^{k}\right)
 = 1+a_{1}(v)\frac{J}{T}+
\left(\frac{a_{1}(v)^2}{2}+a_{2}(v)\right)(\frac{J}{T})^2 \nonumber \\
&& 
\quad + \left(\frac{a_{1}(v)^{3}}{6}+a_{1}(v)a_{2}(v)+a_{3}(v) \right)(\frac{J}{T})^3 
\label{a-ex}  \\ 
&& \quad +\left(\frac{a_{1}(v)^{4}}{24}+\frac{a_{1}(v)^{2}a_{2}(v)}{2}+
\frac{a_{2}(v)^{2}}{2}+a_{1}(v)a_{3}(v)+a_{4}(v) \right)(\frac{J}{T})^4  + \cdots .  
\nonumber
\end{eqnarray}
Substituting (\ref{a-ex}) into (\ref{nlie1}), and 
comparing coefficients of $(J/T)^m$ on both sides, 
we obtain the following integral equation  
for each $m$ ($m \in {\mathbb Z}_{\ge 1}$):  
\begin{eqnarray}
a_{m}(v)=\left( -\frac{h}{J}+\frac{2}{v(v+i)} \right)\delta_{m,1}- \int_{C}\frac{dy}{\pi}
 \frac{\frac{a_{m}(y)}{2}+A_{m}(y)}{1+(v-y)^{2}},
  \label{nlie1-mth}
\end{eqnarray}
where $A_{m}(y)$ is made of $ \{ a_{k}(y) \}_{k=1}^{m-1} $:
\begin{eqnarray}
&& A_{1}(y)=0, \quad 
A_{2}(y)=\frac{a_{1}(y)^{2}}{8}, \quad 
A_{3}(y)=\frac{a_{1}(y)a_{2}(y)}{4}, \nonumber \\
&& A_{4}(y)=\frac{-a_{1}(y)^{4}+24a_{2}(y)^{2}+48a_{1}(y)a_{3}(y)}{192},  \dots .
\end{eqnarray} 
We can solve (\ref{nlie1-mth}) recursively.
The first few terms of $\{a_{m}(v)\}$ are 
\begin{eqnarray}
&& \hspace{-20pt} a_{1}(v)=-\frac{h}{J}-\frac{2 i}{v(1 + v^2)},
\quad 
a_{2}(v)= \frac{h}{J( 1 + v^2 )} + 
    \frac{2 i v}{( 1 + v^2 )^2}, 
\nonumber  \\ 
&& \hspace{-20pt} a_{3}(v)= -\frac{h}{J(1+v^2)},
 \nonumber  \\
&& \hspace{-20pt} 
a_{4}(v)= -\frac{2iv(3+3v^{2}+2v^{4})}{3(1+v^2)^{4}}-
\frac{h(1-3v^{2})}{3J(1+v^2)^{3}}-
\frac{ih^{2}v}{2J^{2}(1+v^2)^{2}}-
\frac{h^{3}}{12J^{3}(1+v^2)}, \nonumber \\
&& \hspace{30pt} \dots . 
\end{eqnarray}
One sees that only $a_{1}(v)$ has a pole at the origin. 
One can also solve the integral equation (\ref{nlie2})
 based on the results of (\ref{nlie1}).
 We assume the following expansion for small $|J/T|$: 
\begin{equation}
G(v,\xi)=\sum_{k=0}^{\infty}g_{k}(v,\xi)\left(\frac{J}{T}\right)^{k}.
 \label{ex-g}
\end{equation}
Substituting (\ref{ex-g}) into (\ref{nlie2}), 
we can obtain the coefficients. The first few terms of 
$\{ g_{k}(v,\xi) \}$ are 
\begin{eqnarray}
&& \hspace{-20pt} g_{0}(v,\xi)=\frac{-i}{( 1 + {( v - \xi ) }^2 ) ( v - \xi ) },
\nonumber \\
&& \hspace{-20pt} g_{1}(v,\xi)= \frac{i(2v -\xi)}
  {( 1 + v^2 ) ( 1 + {( v - \xi ) }^2 ) 
    ( 1 + {\xi}^2 )}
 +\frac{h}{2J(1+(v-\xi)^2)},
  \nonumber \\
&& \hspace{-20pt} g_{2}(v,\xi)= -\frac{i(2v -\xi)\xi^2}
  {( 1 + v^2 ) ( 1 + {( v - \xi ) }^2 ) 
    ( 1 + {\xi}^2 )^{2} } -
\frac{h(2 + 2 v^2 - 2 v \xi + {\xi}^2)}
  {2J( 1 + v^2 ) ( 1 + {( v - \xi ) }^2 ) 
    ( 1 + {\xi}^2 ) } , \nonumber \\
&& \hspace{-20pt} g_{3}(v,\xi)=
-i \bigl(
\left(6 \xi ^2+8\right) v^5-\xi  \left(9 \xi ^2+10\right) v^4+
4 \left(2 \xi ^2+3\right) v^3  \nonumber \\ 
&&  +\xi  \left(3 \xi ^4-4 \xi ^2-9\right)
   v^2+2 \left(4 \xi ^4+9 \xi ^2+6\right) v-\xi  
\left(\xi ^4+3 \xi ^2+3\right)
\bigr) \nonumber \\
&& /\left(3 \left(v^2+1\right)^3 
\left((v-\xi )^2+1\right) \left(\xi
   ^2+1\right)^3\right) \nonumber \\
&& +\frac{h\left(
\left(v^2+1\right) \xi ^4-2 v \left(v^2+2\right) \xi ^3+
\left(2 v^4+7 v^2+1\right) \xi ^2-2 v\xi +3 v^2-1
\right)}{2 J \left(v^2+1\right)^2
   \left((v-\xi )^2+1\right) \left(\xi ^2+1\right)^2}   \nonumber \\
&& +\frac{h^{2}\left(i \xi -2 i v\right)}
{4 J^2 \left(v^2+1\right) \left((v-\xi )^2+1\right) \left(\xi ^2+1\right)}
 -\frac{h^3}{24 J^3 \left((v-\xi )^2+1\right)},
 \dots .
\end{eqnarray}
One see that only $g_{0}(v,\xi)$ has a pole at $v=\xi$. 
Taking note on the above results and 
substituting (\ref{a-ex}) and (\ref{ex-g}) into (\ref{density}), 
we can calculate the coefficients of the HTE for the density matrix 
in a finite magnetic field $h$ only by taking residues at the 
origin. 
Then we can calculate the HTE of the two point functions 
(\ref{2ndzz})-(\ref{2ndpm}) by using the result on the HTE of 
 matrix elements of the density matrix. 
For example, the HTE of a 3rd neighbor correlation function at 
$h=0$ up to the order of 25 is given as follows:
\begin{eqnarray}
&& \hspace{-20pt} <\sigma_{j}^{z}\sigma_{j+3}^{z}>
=-t^{3}-t^{4}+\frac{58t^{5}}{15}+\frac{382t^{6}}{45}-\frac{545t^{7}}{63}
-\frac{14473t^{8}}{315}-\frac{3227t^{9}}{405}
\nonumber \\ 
&& +\frac{2715697t^{10}}{14175
}+\frac{34762571t^{11}}{155925}
-\frac{31700839t^{12}}{51975}
-\frac{1354469786t^{13}}{868725}
\nonumber \\ 
&&
+\frac{405659182t^{14}}{363825}
 +\frac{4993399351951t^{15}}{638512875}
+\frac{1774892665829t^{16}}{638512875} 
\nonumber \\ 
&&
-\frac{337708981437007t^{17}}{10854718875}
-\frac{1373907665967541t^{18}}{32564156625}
+\frac{15824076710853959t^{19}}{168741538875}
\nonumber \\ && 
+\frac{2538703407655267579t^{20}}{9280784638125}
-\frac{1840071379448207089t^{21}}{12993098493375}
\nonumber \\ 
&&
-\frac{945676556978993784629t^{22}}{714620417135625}
-\frac{10941168707397851825299t^{23}}{16436269594119375}
\label{3rdh=0} \\ 
&& +
\frac{251602687993890101833103t^{24}}{49308808782358125}
+
\frac{4135206247498584771500543
   t^{25}}{528308665525265625}
   +\cdots \nonumber ,
\end{eqnarray}
where we put $t=J/T$. 
We also calculated the HTE of $<\sigma_{j}^{z} \sigma_{j+2}^{z}>$ 
for finite magnetic fields $h=2,4,6,8$ 
up to the order of 30; 
$<\sigma_{j}^{z} \sigma_{j+3}^{z}>$ 
for $h=2,4,6,8$ 
up to the order of 20; 
$<\sigma_{j}^{+} \sigma_{j+2}^{-}>$   
for $h=2,4,6,8$ 
up to the order of 30. 
In \cite{FK02,F03}, the HTE of correlation functions for zero magnetic field 
were calculated up to the order of 19 by other method.
It is remarkable that we can also treat nonzero magnetic field case and 
the orders of our HTE are higher than the ones in \cite{FK02,F03}.

\begin{figure}
\psfrag{zz2af}{$<\sigma_{j}^{z} \sigma_{j+2}^{z}>$}
\begin{center}
\includegraphics[width=1\textwidth]
{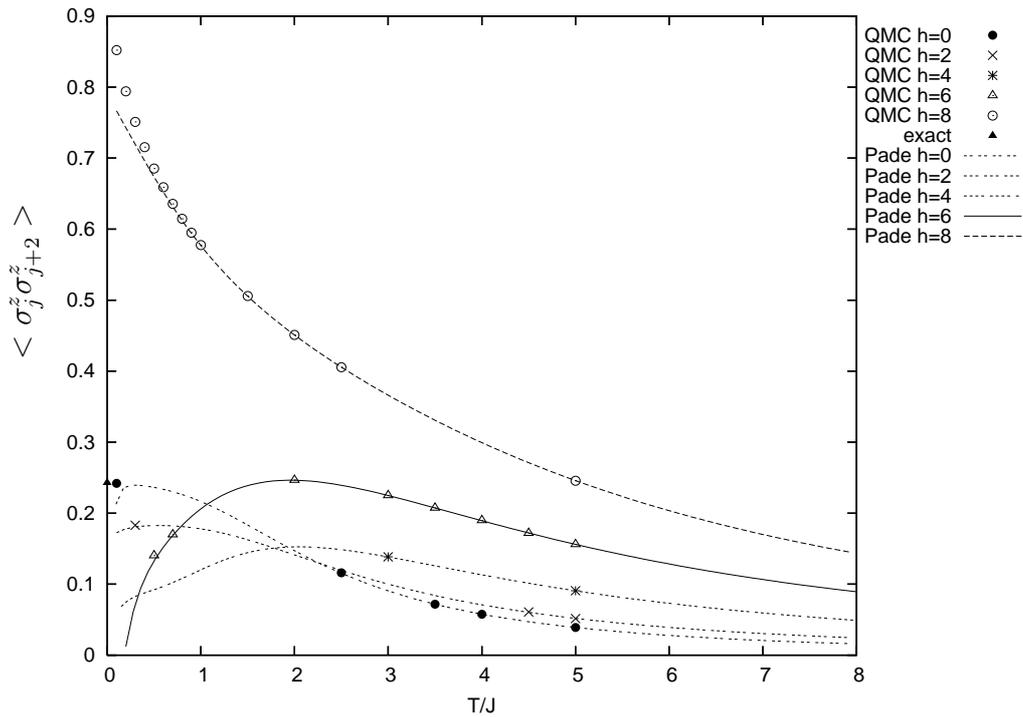}
\end{center}
\caption{Temperature $T$ dependence of 
$<\sigma_{j}^{z} \sigma_{j+2}^{z}>$ for $J>0$ with a magnetic field $h$. 
We have plotted the Pad\'{e} approximations 
of order $[21,21]$ for $h=0$ and $[15,15]$ for $h=2,4,6,8$ 
([$n$,$d$] means that the numerator is a degree $n$ polynomial of $J/T$ and  
 the denominator is a degree $d$ polynomial of $J/T$). 
An exact value of $<\sigma_{j}^{z} \sigma_{j+2}^{z}>$ 
at $(T,h)=(0,0)$ was calculated in \cite{Ta77}.
}
\label{zz2af}
\end{figure}

\begin{figure}
\psfrag{zz3af}{$<\sigma_{j}^{z} \sigma_{j+3}^{z}>$}
\begin{center}
\includegraphics[width=1\textwidth]
{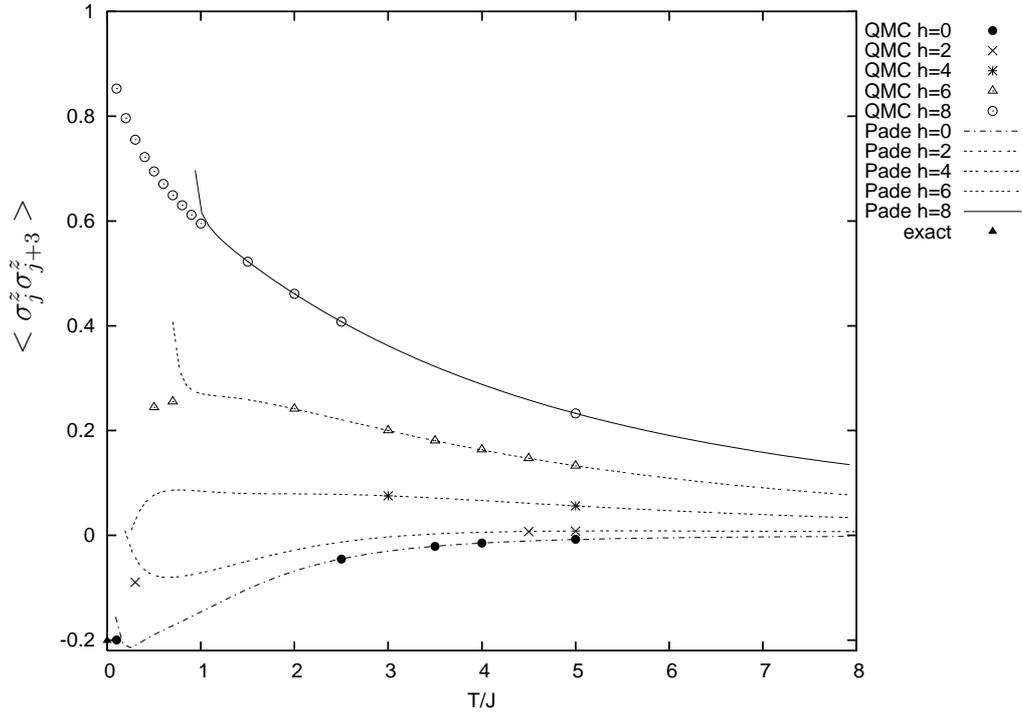}
\end{center}
\caption{Temperature $T$ dependence of 
$<\sigma_{j}^{z} \sigma_{j+3}^{z}>$ for $J>0$ with a magnetic field $h$. 
We have plotted the Pad\'{e} approximations 
of order $[12,13]$ for $h=0$ and $[10,10]$ for $h=2,4,6,8$. 
An exact value of $<S_{j}^{z} S_{j+3}^{z}>=\frac{1}{4}<\sigma_{j}^{z} \sigma_{j+3}^{z}>$ 
at $(T,h)=(0,0)$ was calculated in \cite{SSNT03}.
}
\label{hteAF}
\end{figure}

\begin{figure}
\psfrag{zz3f}{$<\sigma_{j}^{z} \sigma_{j+3}^{z}>$}
\begin{center}
\includegraphics[width=1\textwidth]{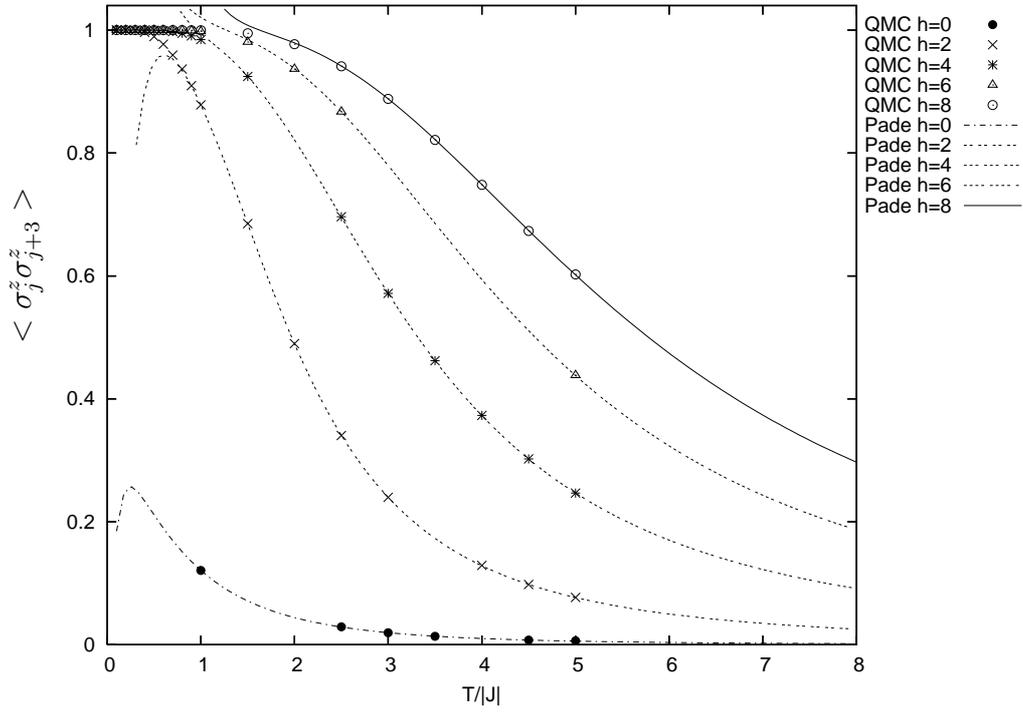}
\end{center}
\caption{Temperature $T$ dependence of 
$<\sigma_{j}^{z} \sigma_{j+3}^{z}>$ for $J<0$ with a magnetic field $h$. 
We have plotted the Pad\'{e} approximations 
of order $[12,13]$ for $h=0$ and $[10,10]$ for $h=2,4,6,8$. 
}
\label{hteF}
\end{figure}

\begin{figure}
\psfrag{greenaf}{$<\sigma_{j}^{+} \sigma_{j+2}^{-}>$}
\begin{center}
\includegraphics[width=1\textwidth]{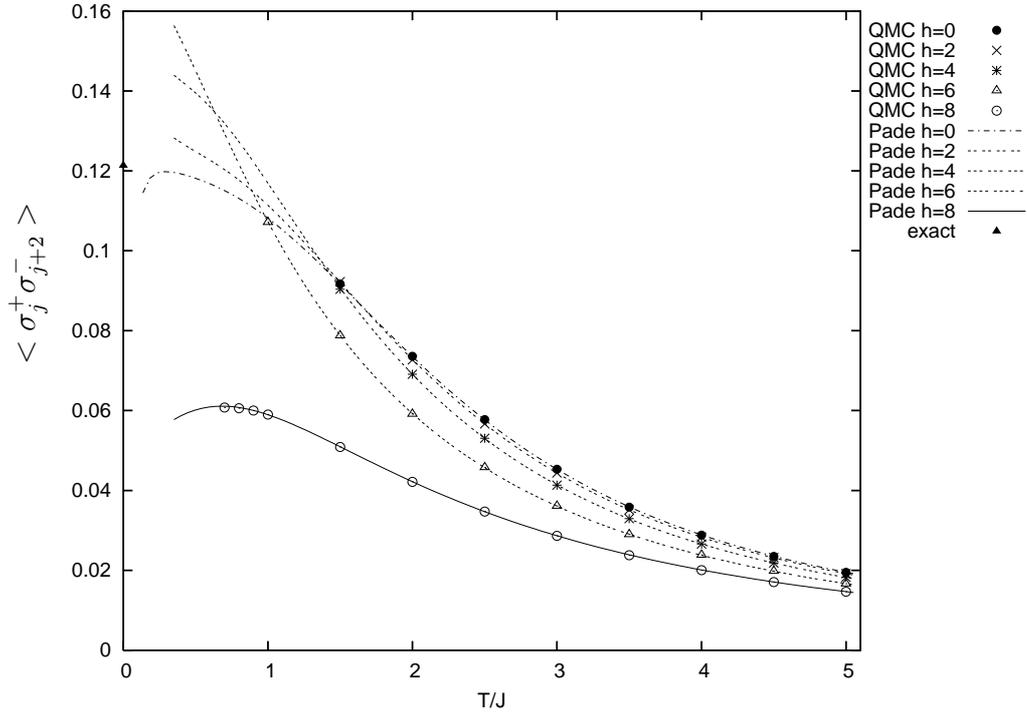}
\end{center}
\caption{Temperature $T$ dependence of 
$<\sigma_{j}^{+} \sigma_{j+2}^{-}>$ for $J>0$ with a magnetic field $h$. 
We have plotted the Pad\'{e} approximations 
of order $[21,21]$ for $h=0$ and $[15,15]$ for $h=2,4,6,8$. 
An exact value of $<\sigma_{j}^{z} \sigma_{j+2}^{z}>
=2<\sigma_{j}^{+} \sigma_{j+2}^{-}>$  
at $(T,h)=(0,0)$ was calculated in \cite{Ta77}.
}
\label{hteGreenAF}
\end{figure}

\begin{figure}
\psfrag{greenf}{$<\sigma_{j}^{+} \sigma_{j+2}^{-}>$}
\begin{center}
\includegraphics[width=1\textwidth]{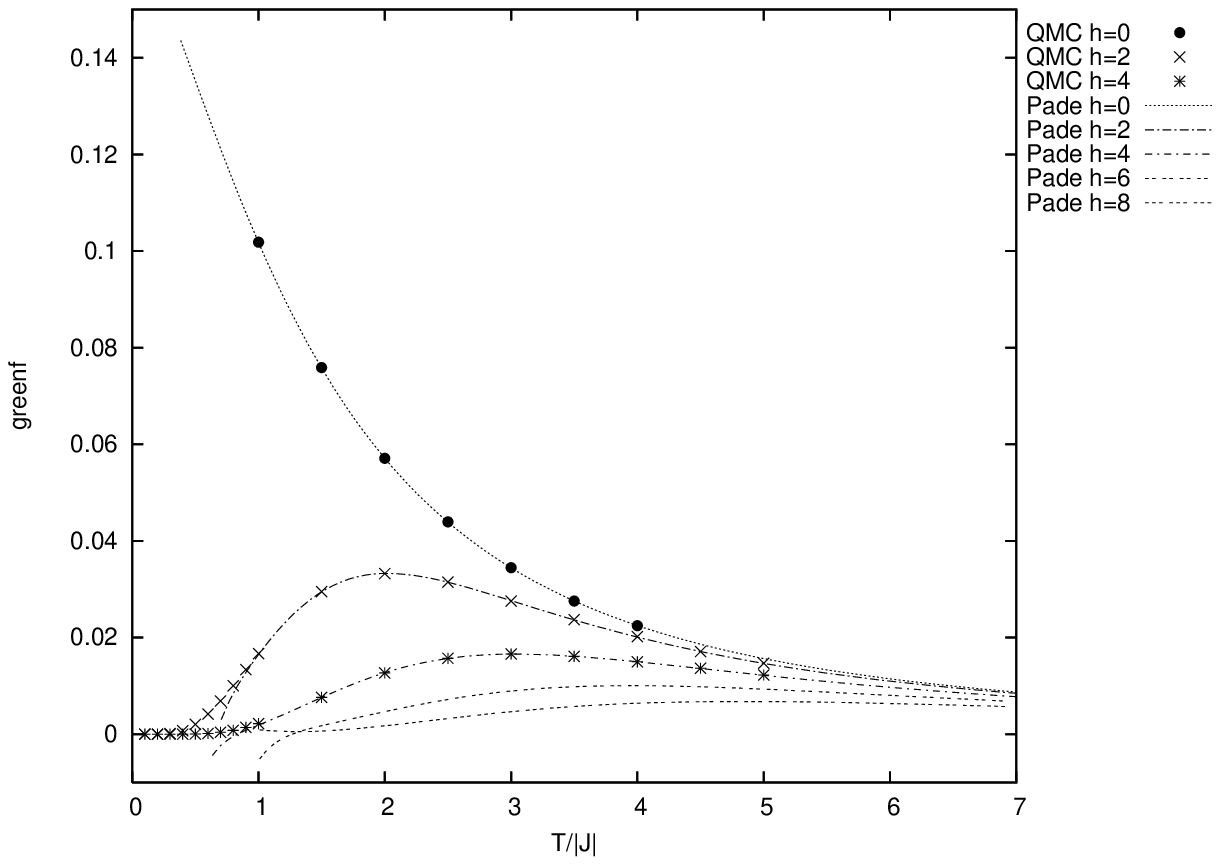}
\end{center}
\caption{Temperature $T$ dependence of 
$<\sigma_{j}^{+} \sigma_{j+2}^{-}>$ for $J<0$ with a magnetic field $h$. 
We have plotted the Pad\'{e} approximations 
of order $[21,21]$ for $h=0$, $[15,15]$ for $h=2,4,6$ 
and $[14,16]$ for $h=8$. 
}
\label{hteGreenF}
\end{figure}

We have plotted the Pad\'{e} approximations of the HTE of these 
correlation functions for finite magnetic fields $h$  
in Figures \ref{zz2af}-\ref{hteGreenF}. 
Once we introduce the HTE, we can also consider $J<0$ case by analytic continuation, 
 though  original multiple integral formula (\ref{density}) was defined for $J>0$ case. 
We have also plotted quantum Monte Carlo simulation (QMC) data by Shiroishi \cite{Sh06} 
based on the open source software in the ALPS project \cite{ALPS,SSE}. 
The system size of the simulations is $L=128$. 
Our HTE results agree well with these QMC data except in the very low temperature regions. 
At least in the region where the Pad\'{e} approximation seems to converge, 
$<\sigma_{j}^{z} \sigma_{j+2}^{z}>$ for $J>0$ at $h=0$ and for $J<0$,  
$<\sigma_{j}^{z} \sigma_{j+3}^{z}>$ for $J<0$ and 
$<\sigma_{j}^{+} \sigma_{j+2}^{-}>$ at $h=0$ 
monotonously decrease with respect to temperature 
(here we omit a figure of $<\sigma_{j}^{z} \sigma_{j+2}^{z}>$ for $J<0$); 
$<\sigma_{j}^{z} \sigma_{j+3}^{z}>$ for $J>0$ at $h=0$ 
monotonously increases with respect to temperature. 
When $h>0$ in figures \ref{zz2af},\ref{hteAF},\ref{hteGreenF}, peaks appear 
at non-trivial temperatures 
. 
\section{Concluding remarks}
In this paper, we have evaluated the multiple integral formula 
of the density matrix by using the HTE method, and thereby obtained the HTE 
of several two point correlation functions. 
Together with our previous paper \cite{TSh05} on $P(m)$, 
we recognized that the HTE method is efficient to 
evaluate the multiple integral formulae of correlation functions \cite{GKS04-3}-\cite{GKS04-2}. 
In \cite{BGKS06}, the multiple integral formula 
(\ref{density}) has been reduced to finite sums over products of 
single integrals for short segments of length 2 and 3.  
If a similar reduction is done for length 4, one will be able to 
calculate the HTE whose order is higher than the one in this paper.  \\
{\bf Acknowledgments} 
We would like to thank Masahiro Shiroishi 
for discussions 
and providing us his unpublished QMC data \cite{Sh06}.
Mathematica Computation in this work was carried out at the 
Yukawa Institute Computer Facility. 
 

\begin{thebibliography}{99}
\bibitem{GKS04-3}
F. G\"ohmann, A. Kl\"umper and A. Seel: 
J. Phys. A: Math. Gen. 38 (2005) 1833-1841; 
cond-mat/0412062.

\bibitem{GKS04-1}
F. G\"ohmann, A. Kl\"umper and A. Seel: 
J. Phys. A: Math. Gen. 37 (2004) 7625--7651; 
hep-th/0405089.

\bibitem{GKS04-2}
F. G\"ohmann, A. Kl\"umper and A. Seel: 
Physica B359-361(2005)807-809; cond-mat/0406611.

\bibitem{GS05}
F. G\"ohmann, A. Seel: 
Theor. Math.Phys. 146 (2006) 119-130; Teor.Mat.Fiz. 146 (2006) 146-160;
hep-th/0505091.

\bibitem{GHS05}
F. G\"ohmann, N.P. Hasenclever, A. Seel:
J.Stat.Mech. 0510 (2005) P015; 
cond-mat/0509765.

\bibitem{S85} 
M. Suzuki: {\it Phys. Rev.} {\bf B31} (1985) 2957--2965.

\bibitem{SI87} 
M. Suzuki and M. Inoue:
{\it Prog. Theor. Phys.} {\bf 78} (1987) 787--799.

\bibitem{K92} 
 A. Kl\"umper: 
{\it Ann. Physik} {\bf 1} (1992) 540--553;
{\it Z. Phys.} B {\bf 91} (1993) 507--519.

\bibitem{DD92}
C. Destri and H. J. de Vega: 
{\it Phys. Rev. Lett.} {\bf 69} (1992) 2313-2317; 
Nucl. Phys. B438[FS](1995)413-454; hep-th/9407117. 

\bibitem{S03}
M. Suzuki: J. Stat. Phys. 110 (2003) 945-956; 
Physica A 321 (2003) 334-339.

\bibitem{JMMN92}
M. Jimbo, K. Miki, T. Miwa and A. Nakayashiki:
 Phys. Lett. A {\bf 168} (1992) 256-263.

\bibitem{JM96}
  M. Jimbo and T. Miwa:
  J. Phys. A: Math. Gen. {\bf 29} (1996) 2923-2958.

\bibitem{KMT00}
N. Kitanine, J. M. Maillet, V. Terras:
Nucl. Phys. B {\bf  567} (2000) 554-582; 
math-ph/9907019.

\bibitem{TSh05} 
Z.Tsuboi, M.Shiroishi: J.Phys.A:Math.Gen.38(2005)L363-L370; cond-mat/0502569. 

\bibitem{BG05}
M. Bortz and F. G\"ohmann: Eur. Phys. J. B 46 (2005) 399-408;
 cond-mat/0504370.

\bibitem{FK02}
N. Fukushima, Y. Kuramoto,  
J. Phys. Soc. Jpn. 71 (2002) 1238-1241; cond-mat/0110550.

\bibitem{F03}
N. Fukushima: J. Stat. Phys. 111 (2003) 1049-1090; cond-mat/0212123.

\bibitem{Sh06}
M. Shiroishi: private communications. 

\bibitem{ALPS}
F. Alet, P. Dayal, A. Grzesik, A. Honecker, M. Koerner, 
A. Laeuchli, S.R. Manmana, I.P. McCulloch, F. Michel,
 R.M. Noack, G. Schmid, U. Schollwoeck, F. Stoeckli, S. Todo, S. Trebst, M. Troyer,
 P. Werner, S. Wessel: 
J. Phys. Soc. Jpn. Suppl. 74 (2005) 30-35; 
cond-mat/0410407. (see also http://alps.comp-phys.org)

\bibitem{SSE}
F. Alet, S. Wessel and M. Troyer:
Phys. Rev. E 71 (2005) 036706; cond-mat/0308495. 

\bibitem{Ta77}
M. Takahashi: J. Phys. C 10 (1977) 1289-1301.

\bibitem{SSNT03}
K. Sakai, M. Shiroishi, Y. Nishiyama and M. Takahashi:
Phys. Rev. E {\bf 67} (2003) 065101(R); cond-mat/0302564.

\bibitem{BGKS06} 
H.E. Boos, F. G\"ohmann, A. Kl\"umper, J. Suzuki:
J.Stat.Mech. 0604 (2006) P001; hep-th/0603064.










 

\end{thebibliography}
\end{document}